\documentstyle[prl,aps,multicol,epsfig]{revtex}

\begin{document}
\title{Quantum Dynamics and Statistics
of\\ Atom-Molecule Bose-Einstein Condensate}
\author{Guang-Ri Jin$^{1}$, Chul Koo Kim$^{1}$, and Kyun
Nahm$^{2}$}
\address{$^{1}$ Institute of Physics and Applied Physics, Yonsei University, Seoul 120-749, Korea}
\address{$^{2}$ Department of Physics, Yonsei University, Wonju 220-710,Korea}
\date{\today}
\maketitle

\begin{abstract}
Based on a two-mode boson model, we study nonclassical properties
of the atom-molecule Bose-Einstein condensate. The effects of
nonlinear collisions on the dynamics of the molecular formation is
studied both in classical and quantum treatments. We find that the
conversion from atoms to molecules can be suppressed strongly due
to nonlinearity-induced localization of the atomic population. In
addition, we study statistical properties of the atom-molecule
condensed system by calculating the intensity correlation
functions numerically. We find that the effect of nonlinearity
leads to the appearance of superchaotic molecular pulses, while
maintaining the atomic field sub-Poissonian. The joint quantum
statistical properties of the atoms and the molecules always show
anti-bunching. \newline {PACS numbers: 03.75.-b,05.30.Jp}
\end{abstract}

\begin{multicols}{2}

Recently, possibility of preparing atom-molecule Bose-Einstein
condensate (AMBEC) has attracted wide attentions
\cite{drum,timm,abee,Wynar}. Condensed bosonic atoms can be
converted to a molecular condensate by using either the
photoassociation process \cite{drum,Java99,hein} or the so-called
Feshbach resonance method \cite{timm,abee,Donley,Claussen}.
Wieman's group in JILA \cite{Donley,Claussen} measured the
first-order temporal mutual coherence of atoms and molecules by
using two-photon Ramsey experiment. Similarly with previous works
\cite{Hall,Minardi}, almost coherent Rabi oscillations (Ramsey
interference pattern) were demonstrated. Theoretically, their
experimental results were reproduced partially by using coupled
Gross-Pitaevski (GP) equations
\cite{holland,Kempen,Burnett,Stoof}. According to the mean-field
theory (MFT), large-amplitude coherent oscillation between the
two-field modes is expected. Vardi {\it et al.} \cite{Vardi} found
that quantum-field solutions modify the Bose-enhanced factor of
the oscillation frequency. Moreover, many-body quantum effects
lead to the appearance of collapse and revival of the coherent
oscillations \cite{Java99,Vardi}.

In this letter, we study nonclassical properties of the AMBEC
based on a two-mode bosonic model \cite{Java99,Vardi}. The role of
nonlinearity on quantum dynamics of the AMBEC is studied. We find
that nonzero interspecies and intraspecies interactions result in
modulational instability, and modify the dynamics of the AMBEC.
The conversion of atoms into molecules is shown to be dramatically
suppressed due to nonlinearity-induced localization of the atomic
population. We also study the statistical properties of the AMBEC
by calculating the intensity correlation functions numerically.
Our results show that due to the nonlinearity, the initial
sub-Poissonian molecular field is transformed into superchaotic
molecular pulses, whereas the atomic field remains in the
sub-Poissonian region. In addition, the joint quantum statistical
properties of the atoms and the molecules always exhibit
anti-bunching.

We consider the atom-molecule condensate system coupled via
one-color photoassociation or Feshbach resonance. The total system
can be described phenomenologically by a two-mode bosonic
Hamiltonian \cite{Java99,Vardi} ($\hbar =1$):
\begin{eqnarray}
\hat{H} &=&-\frac{\delta }{2}\hat{a}^{\dagger
}\hat{a}+\frac{g}{2}(\hat{b} ^{\dagger }\hat{a}^{2}+H.c.)+\lambda
_{ab}\hat{a}^{\dagger }\hat{a}\hat{b}
^{\dagger }\hat{b}  \nonumber \\
&&+\frac{\lambda _{a}}{2}\hat{a}^{\dagger }\hat{a}^{\dagger
}\hat{a}\hat{a}+ \frac{\lambda _{b}}{2}\hat{b}^{\dagger
}\hat{b}^{\dagger }\hat{b}\hat{b}, \label{HNL}
\end{eqnarray}%
where $\delta $ denotes the detuning between the molecular and
atomic field. The atomic- and molecular-field operators, $\hat{a}$
and $\hat{b}$ obey the standard bosonic commutation relationship.
The atom-atom, molecule-molecule, and atom-molecule elastic
interactions are described by the s-wave scattering strengthes
$\lambda_{a}$, $\lambda _{b}$, and $\lambda _{ab}$ ($\lambda
_{ba}$) \cite{timm,hein}, respectively. It should be mentioned
that $\lambda _{b}$ and $\lambda _{ab}$ are still unknown
\cite{Wynar,hein}. Following Refs. \cite{Kivshar,Olsen}, we take
$\lambda_b\sim 2\lambda_a$ ($\sim 10^{-3}g$) and
$\lambda_{ab}\sim-1.5\lambda_a$. Due to the conserved particle
number $\hat{N}=\hat{n}_{a}+2\hat{n}_{b}$, where
$\hat{n}_{a}=\hat{a }^{\dagger }\hat{a}$ and
$\hat{n}_{b}=\hat{b}^{\dagger }\hat{b}$, the Hilbert space is
spanned by the basis \cite{Jex92}: {$|\phi _{n}\rangle
=|N-2n\rangle _{a}|n\rangle _{b}$} for $n=0\text{, }1\text{,...,
}[N/2]$, where $[z]$ denotes taking the greatest integer of any
real number $z$. The dynamics of the AMBEC for arbitrary initial
states can be solved by using the Runge-Kutta numerical scheme.

Before investigating the quantum dynamics of the AMBEC, we first
derive the classical equations of motion and discuss their steady
states. The GP-type equations for the atomic and molecular fields
can be obtained from the Heisenberg equations by replacing
$\hat{a}\rightarrow \langle \hat{a}\rangle
=\sqrt{\langle N\rangle }\Phi _{a}$ and $\hat{b}\rightarrow \langle \hat{b}%
\rangle =\sqrt{\langle N\rangle }\Phi _{b}$, where the wave functions for
the two-field modes $\Phi _{j}=|\Phi _{j}|e^{i\phi _{j}}$ ($j=a$, $b$) obey
the normalized condition: $\left\vert \Phi _{a}\right\vert ^{2}+2|\Phi
_{b}|^{2}=1$. To get the steady states and the modulational instability
conditions \cite{Kostrun}, we introduce two conjugate variables \cite%
{Smerzi,YWu}: $x=\left\vert \Phi_{a}\right\vert ^{2}$ and $\varphi
=2\phi _{a}-\phi _{b}$. They obey the canonical relations:
$dx/d\tau=-\partial {\cal {H}}/\partial \varphi$,
$d\varphi/d\tau=\partial {\cal {H}}/\partial x$, and the canonical
Hamiltonian is given by
\begin{eqnarray}
{\cal H}=\Delta x-\frac{\Lambda }{2}x^{2}-x\sqrt{1-x}\cos
(\varphi),
\end{eqnarray}%
where the scaled time $\tau =\sqrt{2}Gt$. In derivation of the
above equations, we have introduced
\begin{eqnarray}
\Delta &=&(\sqrt{2}G)^{-1}\left( \delta -\Lambda _{ab}+\Lambda
_{b}/2\right) ,  \label{delta} \\
\Lambda &=&(G/\sqrt{2})^{-1}\left( \Lambda _{a}-\Lambda
_{ab}+\Lambda _{b}/4\right) ,  \label{Lam}
\end{eqnarray}%
where $G=\sqrt{\langle N\rangle }g$ and $\Lambda _{ij}=\langle
N\rangle \lambda _{ij}$. The stationary solutions obey $\left(
dx/d\tau \right) _{x_{0}}=\left( d\varphi /d\tau \right) _{\varphi
_{0}}\equiv 0$. From the relation $x_{0}\sqrt{1-x_{0}}\sin
(\varphi _{0})=0$, we find that there are three kinds of fixed
points: (i) pure molecular phase with $x_{0}=0$, (ii) in-phase
type of steady state with $\varphi _{0}=0$, (iii) out-of-phase
type of steady state with $\varphi _{0}=\pi $ \cite{Kivshar}. The
value of $x_{0}$ determined by a cubic equation
\begin{equation}
\Delta -\Lambda x_{0}-\frac{1-3x_{0}/2}{\sqrt{1-x_{0}}}\cos
(\varphi _{0})=0 \label{x0}
\end{equation}%
can be solved numerically. It was shown that the atomic phase at
$|\Delta|\rightarrow\infty$ can be converted to the molecular
phase or vice versa by varying the detuning $\Delta$ adiabatically
\cite{Java99,YWu}. In the intermediate region, there are two-type
atom-molecule coexisting phases: the in-phase steady state for
$\varphi _{0}=0$ in the parameter region $\Delta<1$, and the
out-of-phase one for $\varphi _{0}=\pi$ in $\Delta>-1$ region.

Stability analysis of the steady states can be performed by
introducing small perturbations: $x=x_{0}+\delta x$ and $\varphi
=\varphi _{0}+\delta \varphi $. A coupled linearized equation for
the perturbations gives the eigenfrequencies
\begin{equation}
\omega ^{2}=\omega _{0}^{2}-\Lambda x_{0}\sqrt{1-x_{0}}\cos (\varphi _{0}),
\end{equation}%
where $\omega _{0}=\sqrt{x_{0}(1-3x_{0}/4)\cos ^{2}(\varphi
_{0})/(1-x_{0})}$. For a negligible nonlinear interaction $\Lambda
=0$, aroused from either $\Lambda_{i}=\Lambda_{ij}=0$ or $\Lambda
_{ab}=\Lambda _{a}+\Lambda _{b}/4$, both of the two steady states
($\varphi _{0}=0$, $\pi $) are stable \cite{YWu}. However, for
nonzero $\Lambda $, the above two steady states can be
modulationally unstable when the eigenfrequencies become imaginary
\cite{Agrawal}, i.e., $\omega^{2}<0$. In this case, the
perturbations in the AMBEC undergo exponential growth with the
growth rate being the imaginary part of $\omega$. We find that,
for $\varphi _{0}=0$, the MI occurs in the positive $\Lambda $
region with $\Lambda >1$ and $\Delta <1$. For $\varphi _{0}=\pi $,
the AMBEC is unstable in the negative $\Lambda$ region with $
\Lambda <-1$ and $\Delta >-1$. The MI region (and also the
dynamics) of the AMBEC is invariant under the transformation
$\Lambda \rightarrow -\Lambda $, $\varphi \rightarrow \pi -\varphi
$, and $\Delta \rightarrow -\Delta $. Thus, in the following we
only consider the case of positive $\Lambda$.

Dynamical evolution of the AMBEC can be described by the fraction
of atoms converted to molecules \cite{Java99,Kostrun},
$f_{M}=2\langle \hat{b}^{\dagger }\hat{b}\rangle/\langle
\hat{N}\rangle$, which measures the ratio of the atom number in
the bound state to the total atom number. In the MFT,
$f_{M}=2|\Phi _{b}|^{2}$. For the negligible nonlinear interaction
$\Lambda =0$, there are analytical solutions for $f_{M} $
\cite{Ish}. Considering the initial condition $x(0)=1$, then
$f_{M}=2f_{1} \text{sn}^{2}(\sqrt{f_{2}}\tau /\sqrt{2}\text{,
}m)$, where sn($z$, $m$) is a Jacobi elliptic function with the
elliptic modulus $m=f_{1}/f_{2}$, and $f_{1,2}=\frac{1}{4}[\Delta
^{2}+2\mp \sqrt{\Delta ^{2}(\Delta ^{2}+4)}]$. In particular, for
the exact ``resonant" case with $\Delta =0$, $f_{M}=\tanh
^{2}\left( \tau /2\right) $ since $m=1$. Starting from the zero
initial value, $f_{M}$ grows quickly to the maximum value $1$
without any oscillation (the dotted line in Fig. 1 (a)), which
corresponds to a complete and irreversible molecule formation
process. Nonzero $\Delta$, however, results in periodic
oscillations of $f_{M}$ with smaller amplitude. If $\Delta >\Delta
_{c}\equiv 1/\sqrt{2}$, the amplitude of $f_{M}$ is always less
than $1/2$.

To investigate the effect of nonzero $\Lambda$, we need to know
the values of the two-body interaction strengthes. However
$\lambda_b$ and $\lambda_{ab}$ have not been determined even in
low-energy region \cite{Wynar,hein}. Following Ref. \cite{Olsen},
we adopt $\lambda_b=2\lambda_a$ and $\lambda_{ab}=-1.5\lambda_a$.
Nonzero $\Lambda$ can be achieved by tuning $\lambda_a$, such as
$\lambda_a=1.0\times 10^{-3}g$ ($\Lambda=0.4243$),
$\lambda_a=2.22\times 10^{-3}g$ ($\Lambda=0.9419$), and
$\lambda_a=4.0\times 10^{-3}g$ ($\Lambda=1.6971$). All the above
parameters are in the stable region. The MFT results show that the
effect of nonlinearity leads to periodic oscillations between
atoms and molecules. As shown in Fig. 1 (a), the amplitude of
$f_{M}$ decreases monotonically with the increase of $\Lambda$. If
$\Lambda>0.94$, the whole curve of $f_{M}$ is below $1/2$, {\it
i.e.}, no more than 50 percent of atoms are converted to
molecules. This nonlinearity-induced localization of the atomic
population is similar to the macroscopic self-trapping in the
two-component BEC \cite{Smerzi}. However, unlike the two-mode BEC
system with Josephson-like coupling, the oscillation amplitude in
AMBEC reduces continuously from unity to zero.

In Fig. 1 (b), we evaluate numerically the quantum solutions of
$f_{M}$ for an initial number state $\left\vert \psi
(0)\right\rangle =\vert N\rangle _{a}\vert 0\rangle_{b}$. In the
absence of nonlinearity $\Lambda=0$ and $\Delta=0$, the quantum
solution of $f_M$ (the dotted line in Fig. 1 (b)) breaks away from
the MFT result around the point $\tau\sim7.03$, where all the
particles are in molecular state. This departure can not be
explained in the framework of the MFT. A semiclassical analysis of
the AMBEC in the vicinity of the all-molecules phase shows that
there are two contributions to the atomic occupation \cite{Vardi}:
(i) spontaneous dissociation into atomic vacuum, which is
proportional to $\sqrt{N}\sinh^2(\tau/2)$, (ii) stimulated part
proportional to $\langle \hat{n}_a\rangle$, which is negligible
near the all-molecules phase. The Bose-enhanced spontaneous
dissociation into the atomic vacuum results in collapse and
revival of the atom-molecular oscillations \cite{Java99,Vardi},
which is just similar with the phenomenon observed in the
two-photon down-conversion process \cite{Jex92}.

We further consider quantum solution of $f_M$ for the case of
$\Lambda\neq 0$. As shown in Fig. 1 (b), we find that the quantum
results almost coincide with the corresponding classical solutions
shown in Fig. 1 (a). Both of them exhibit ``periodic" oscillations
with decreased amplitude. The agreement of two solutions
originates from the fact that the trajectories do not reach the
all-molecules phase, and the contribution of quantum fluctuation
of atomic mode is {\it not} significant compared to the coherent
stimulated process. In Fig. 1 (c), we compare the quantum solution
(red solid-line) with the MFT solution (dotted line) for
$\Lambda=0.4243$. We find that the quantum solution is in fact the
collapsed oscillations. From Fig. 1(b), we also find that the
collapse time is enhanced significantly by the nonlinearity, {\it
i.e.}, the coherent oscillations are damped more slowly with the
increase of $\Lambda$.

%%%%%%%%%%%%%%%%%%%%%%%Fig. 1%%%%%%%%%%%%%%%%%%%%%%%%%%%%%%%%
\vskip -0.2cm
\begin{figure}
\begin{center}
\epsfxsize=8cm\epsffile{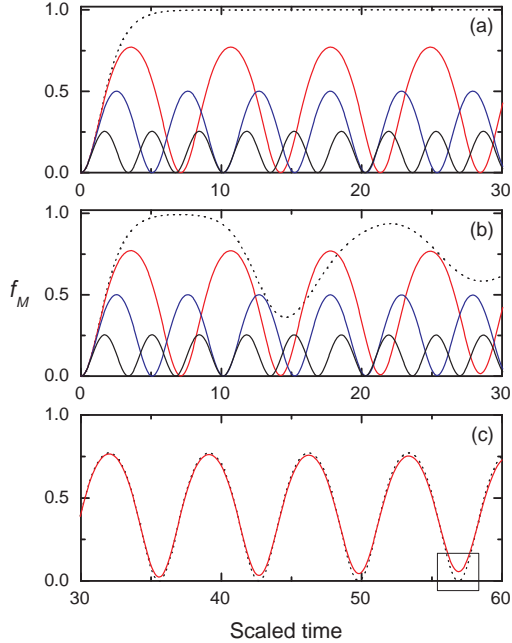} \caption{(Color online).
$f_M$ versus the scaled time $\tau$ for (a) mean-field solutions,
(b) quantum solutions, and (c) comparison of the two predictions
for $\Lambda=0.4243$. The parameters are $\Delta=0$, $N=10^4$, and
the nonlinearity in (a) and (b) are: $\Lambda=0.0$ (the dotted
lines), $\Lambda=0.4243$ (the red lines), $\Lambda=0.9419$ (the
blue lines), and $\Lambda=1.6971$ (the black solid lines).}
\end{center}
\end{figure}
%%%%%%%%%%%%%%%%%%%%%%%Fig. 1%%%%%%%%%%%%%%%%%%%%%%%%%%%%%%%%

The observation of the first-order temporal coherence between
atoms and molecules opens up the possibility to form molecular
condensate. However, there remains many open questions to be
answered \cite{Donley}, such as the detailed quantum state of the
molecules, and their higher-order coherence functions in the
temporal domain. Recently, Meiser {\it et al.} \cite{Meystre}
studied the short-time dynamics of molecule formation and the
second-order correlation function by using perturbation theory.
They found that the second-order degree of coherence of the
molecular field for an initial number state approaches unity in
the limit of large $N$. Motivated by their work, we calculate the
equal-time intensity correlation functions beyond the short-time
limit:
\begin{eqnarray}
g_{j}^{(2)}=\frac{\langle \hat{n}_{j}(\hat{n}_{j}-1)\rangle }{
\left\langle \hat{n}_{j}\right\rangle ^{2}}.  \label{g2j}
\end{eqnarray}
It is known from quantum optics that if $0\leq g_j^{(2)}<1$ then
the field is in the so-called sub-Poissonian. The fields in a
sub-Poisson distribution always exhibit nonclassical {\it
anti-bunching} phenomena. $g_j^{(2)}=1$ represents a coherent
state with Poisson distribution, while $g_j^{(2)}>1$ is
characteristic for a super-Poisson distribution. In particular,
for thermal or chaotic field, $g_j^{(2)}=2$.

%%%%%%%%%%%%%%%%%%%%%%%%Fig. 2%%%%%%%%%%%%%%%%%%%%%%%%%%%%%%%%
\vskip -0.5cm
\begin{figure}[htbp]
\begin{center}
\epsfxsize=8cm \epsffile{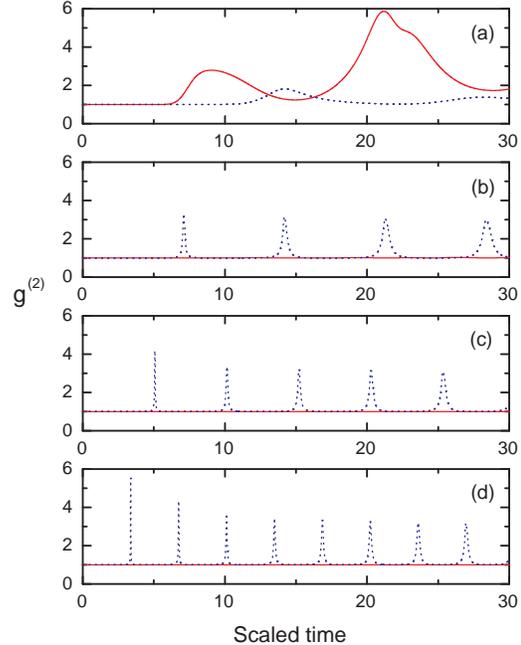} \caption{(Color online).
$g_a^{(2)}$ and $g_b^{(2)}$ versus the scaled time $\tau$ for
various $\Lambda$. From top to bottom: (a) $\Lambda=0$, (b)
$\Lambda=0.4243$, (c) $\Lambda=0.9419$, and (d) $\Lambda=1.6971$.
The red solid lines are for $g_a^{(2)}$ and the blue dotted lines
for $g_b^{(2)}$. Other parameters are the same with Fig. 2.}
\end{center}
\end{figure}
%%%%%%%%%%%%%%%%%%%%%%%%%%%%%%%%%%%%%%%%%%%%%%%%%%%%%%%%%%%%%%%%%%%

Starting from the number state, the initial stage of the
second-order coherence functions of two field modes obey
$g_a^{(2)}=1-1/N$ and $g_b^{(2)}=1-4/N+{\cal O}(N^{-2})$
\cite{Meystre}. Both of them are less than $1$. For zero
nonlinearity $\Lambda=0$, as shown in Fig. 2(a), we find that the
two field modes are transformed from sub-Poissonian to
super-Poissonian, successively. Similarly with previous
discussions, the AMBEC system with zero nonlinearity shares the
same model of degenerate parametric amplifier \cite{walls}.
Quantum fluctuations of the atomic vacuum leads to the appearance
of super-Poissonian, and the peaks of $g_a^{(2)}$ near the points
$\tau\sim 7.03$ and $21.0$, where the particles are almost in the
molecular mode. The magnitude of $g_a^{(2)}>2$ indicates that the
atoms generated from the photodissociation exhibits superchaotic
(strong bunching) behavior \cite{Molmer}.

The second-order coherence function $g_j^{(2)}$ for nonzero
$\Lambda$ are calculated in Fig. 2 (b), (c) and (d).  We find that
the effects of nonlinearity result in sharp peaks in $g_b^{(2)}$
with their positions at the minima of $f_M$. Their magnitudes are
larger than $2$ but not divergent due to the nonzero minimums of
the mean molecular number. Our results imply that the initial
sub-Poissonian statistics of $g_b^{(2)}$ is transformed into a
super-Poissonian, and thereby a superchaotic molecular pulse can
be realized at certain values of time. The second-order coherence
functions of the atomic field are also plotted in Fig. 2 (the red
lines). We find that super-Poissonian statistics of $g_a^{(2)}$ is
fully suppressed for $\Lambda\neq 0$. More specifically, the
intensity correlation function of the atomic field oscillates
between sub- and super-Poissonian with very small amplitude.
Further increase of $\Lambda$, such as to $1.6971$, $g_a^{(2)}<1$,
which implies that the atomic field is always in sub-Poissonian.
The reduction of the super-Poissonian of atomic field to
sub-Poissonian comes also from the suppression of spontaneous
atomic emission. The increase of $\Lambda$ dominates the
contribution of the coherent stimulated process.

Finally, we study the joint quantum statistical properties of the
atoms and the molecules, defined by
\begin{equation}
g_{ab}^{(2)}=g_{ba}^{(2)}\equiv \frac{\langle \hat{n}_{a}\hat{n}%
_{b}\rangle }{\langle \hat{n}_{a}\rangle \langle
\hat{n}_{b}\rangle }.
\end{equation}
We find that $g_{ab}^{(2)}(\tau)$ is always less than $1$.
Physically, such a result means that the atoms and the molecules
do not tend to be created simultaneously, {\it i.e.}, the
anti-bunching between the two field modes. Similar results have
been addressed for optical field in the two-photon down-conversion
process \cite{Jex92}.

In summary, we have studied the role of nonlinearity on quantum
dynamics and statistical properties of the AMBEC. The former
problem is related to the atom-molecule conversion rate. We have
shown that even in the simplest two-mode model without ``rogue
photodissociation" \cite{Javan02}, the conversion can be strongly
settled down due to the nonlinear analog of macroscopic
self-trapping effect. We also studies statistical properties of
the AMBEC by calculating the intensity correlation functions
numerically. Our results show that the nonlinearity $\Lambda$ in
the stable region can modify both $g_{a}^{(2)}$ and $g_{b}^{(2)}$
significantly. The initial sub-Poissonian molecular field is
transformed into superchaotic molecular pulses, whereas the atomic
field remains in sub-Poissonian region. Our result also show that
the joint quantum statistical properties of the atoms and the
molecules always exhibit anti-bunching.

It should be mentioned that Prataviera {\it et al.}
\cite{Prataviera} have studied higher-order mutual coherence of
two chemically different species: the light field and the matter
field. Their scheme make use of the combination of the traditional
absorption photodetector and a matter-wave detector relied on
photoionization of the atoms. Similarly, photoionization of the
diatomic molecules can be proposed in the atom-molecule
photoassociation experiments to measure the molecules and their
statistical properties directly.

This work is supported in part by the BK21 and by Korea Research
Foundation (KRF-2003-005-C00011). We would like to express our
sincere thanks to Dr. Guo-Hui Ding for helpful discussions.

\end{multicols}

\end{document}